\documentclass[twocolumn,showpacs,preprintnumbers,amsmath,amssymb]{revtex4}
\usepackage{graphicx}% Include figure files
\usepackage{dcolumn}% Align table columns on decimal point

\begin{document}
\title{Driven degenerate three-level cascade laser}

\author{Sintayehu Tesfa}
 %\altaffiliation[Present address ]{Physics Department, Debub University, Dilla College, P. O. Box 419.}%Lines break automatically or can be forced with \\
 %\email{sint_tesfa@yahoo.com}
\affiliation{Physics Department, Addis Ababa University, P. O. Box 1176, Addis Ababa, Ethiopia}%

%\affiliation{

\date{\today}

\begin{abstract}We analyze a degenerate three-level cascade
laser coupled to an external coherent light via one of the coupler
mirrors and vacuum reservoir in the other, employing the
stochastic differential equation associated with the normal
ordering. We study the squeezing properties and also calculate the
mean photon number of the cavity radiation. It turns out that the
generated light exhibits up to 98.3\% squeezing under certain
conditions pertaining to the initial preparation of the
superposition and the amplitude of the driving radiation.
Moreover, the mean photon number is found to be large where there
is a better squeezing. Hence it is believed that the system under
consideration can generate an intense squeezed light.
\end{abstract} 
 \maketitle

 \section{INTRODUCTION}

Interaction of three-level atoms with a radiation has attracted a
great deal of interest in recent years
\cite{1,2,3,4,5,6,7,8,9,10,11,12}. It is believed that an atomic
coherence is found to be responsible for various important quantum
features of the emitted light. In general, the atomic coherence
can be induced in a three-level atom by coupling the levels
between which a direct transition is dipole forbidden by an
external radiation \cite{1,2,3,4,5,6} or by preparing the atom
initially in a coherent superposition of these two levels
\cite{7,8,9}. It is found that the cavity radiation exhibits
squeezing under certain conditions for both cases
\cite{4,9,10,11}.

In a cascade three-level atom  the top, intermediate, and bottom
levels are conveniently denoted by $|a\rangle$, $|b\rangle$, and
$|c\rangle$ in which a direct transition between levels
$|a\rangle$ and $|c\rangle$ is dipole forbidden. When the
three-level cascade atom decays from $|a\rangle$ to $|c\rangle$
via the level $|b\rangle$ two photons are generated. If the two
photons have identical frequency, then the three-level atom is
referred to as a degenerate. Hence we define a degenerate
three-level laser as a quantum optical system in which degenerate
three-level atoms in a cascade configuration and initially
prepared in a coherent superposition of the top and bottom levels
are injected at a constant rate into a cavity. These atoms are
removed from the cavity after some time. We hence realize that, a
degenerate three-level laser is a two photon device in which
squeezing properties are expected to occur due to the correlation
between these two photons \cite{2,4}.

We consider a degenerate three-level cascade laser coupled to  a
vacuum reservoir via a single-port mirror and the bottom level of
the atoms on the other hand is coupled to the top level by an
external resonant coherent light as shown in Fig. 1. Some
authors have already studied such a scheme in which the atomic
coherence is induced by an external radiation and when
initially the atoms are prepared in the top level \cite{4,10} and
bottom level \cite{6}. They found that the three-level laser in
these cases resemble the parametric oscillator for a strong
radiation. Moreover, recently Saavedra {it{et al}}. \cite{1} studied the
$\lambda$ three-level laser when the atoms are initially prepared
in a coherent superposition and the forbidden transition is induced
by driving with strong external radiation. They found that there
is lasing without population inversion with the favorable noise
reduction occurs for equal population of the two levels and when
the initial coherence is maximum. Therefore, we consider the case
when the atoms are initially prepared to be in an arbitrary atomic
superposition and in addition driven on resonance externally by a
coherent radiation.  We restrict our analysis to the regime of
lasing without population inversion.

In this communication, we study the squeezing properties of the
cavity radiation and we also calculate the mean photon number
using the stochastic differential equation associated with the
normal ordering. We prefer to employ the classical stochastic
relations to the corresponding quantum operator for obvious reason
that it is mathematically easier to deal with. In particular, we
calculate the quadrature variance and the mean photon number for
cases when the atoms are initially prepared to be in the bottom
level and when they are having equal probability of being in the
top and bottom levels.
\begin{picture}(250,160)(-30,-60)
\put (0,55){\line(1,0){80}} \put (0,55.5){\line(1,0){80}} \put
(83,52.5){$|a\rangle$} \put (-20,20){$\Omega$}\put
(80,30){$\hat{a}$}\put (80,-20){$\hat{a}$}\put
(40,55){\vector(0,-1){50}} \put (20,5){\vector(0,1){50}}\put
(-30,15){\vector(1,0){50}}\put (40,25){\vector(1,0){50}}\put
(44,30){$\omega_{a}$} \put (0,5){\line(1,0){80}} \put
(0,5.5){\line(1,0){80}} \put (83,2.5){$|b\rangle$} \put
(0,-45){\line(1,0){80}} \put (0,-45.5){\line(1,0){80}} \put
(83,-47.5){$|c\rangle$} \put (40,5){\vector(0,-1){50}} \put
(20,-45){\vector(0,1){100}}\put (40,-25){\vector(1,0){50}}\put
(44,-17.5){$\omega_{a}$}
\end{picture}
{ FIG. 1:}~{\footnotesize Schematic representation of a coherently
driven degenerate three-level atom in a cascade configuration. The
transitions between $|a\rangle-|b\rangle$ and
$|b\rangle-|c\rangle$ at frequency $\omega_{a}$ each are taken to
be resonant with the cavity. The transition $|b\rangle-|c\rangle$
is dipole forbidden and can be induced by driving the atom
externally with resonant radiation of frequency $2\omega_{a}$. }

\section{ Master equation}

 The interaction of a degenerate three-level atom with a single-mode light can be
 described in the rotating-wave approximation and in the interaction
picture by the Hamiltonian of the form
\begin{align}\label{dt01}\hat{H}_{AR}
=ig\big[\hat{a}(|a\rangle\langle b| + |b\rangle\langle
c|)-(|b\rangle\langle a| + |c\rangle\langle
b|)\hat{a}^{\dagger}\big],\end{align} where $g$ is the coupling
constant, which is taken to be the same for both transitions, and
$\hat{a}$ is the annihilation operator for the cavity mode. On the
other hand, the three-level cascade atom for which its bottom
level is coupled to the top level by a resonant coherent light can
be expressed in the rotating-wave approximation and in the
interaction picture by the Hamiltonian of the form
\begin{align}\label{dt02}\hat{H}_{C} =i{\Omega\over2}\big[|c\rangle\langle a|-|a\rangle\langle
c|\big],\end{align} where $\Omega$ is a real-positive constant
proportional to the amplitude of the coherent driving radiation.
Hence on the basis of Eqs. \eqref{dt01} and \eqref{dt02} the
interaction of a coherently driven three-level atom with the
cavity radiation can be represented in the rotating-wave
approximation and in the interaction picture by the Hamiltonian,
\begin{align}\label{dt03}\hat{H}_{AR} &=ig\big[\hat{a}(|a\rangle\langle b|+ |b\rangle\langle c|)
-(|b\rangle\langle a|+|c\rangle\langle b|)\hat{a}^{\dagger}\big]
\notag\\&+ i{\Omega\over2}\big[|c\rangle\langle
a|-|a\rangle\langle c|\big].\end{align}

In this communication, we take the initial state of a three-level
atom to be
\begin{align}\label{dt04}|\Phi_{A}(0)\rangle\; = C_{a}(0)|a\rangle +
C_{c}(0)|c\rangle,\end{align} where $C_{a}(0) = \langle
a|\Phi_{A}(0)\rangle$ and $C_{c}(0) = \langle
c|\Phi_{A}(0)\rangle$ are probability amplitudes for the atom to
be in the top and bottom levels, respectively. This corresponds to
the fact that the three-level atom is initially prepared to be in
a coherent superposition of the top and bottom levels. Hence the
initial density operator for the atom described by the quantum
state \eqref{dt04} would be
\begin{align}\label{dt05}\hat{\rho}_{A}(0)=\rho_{aa}^{(0)}|a\rangle\langle
a|+\rho_{ac}^{(0)}|a\rangle\langle
c|+\rho_{ca}^{(0)}|c\rangle\langle
a|+\rho_{cc}^{(0)}|c\rangle\langle c|,\end{align} where
\begin{align}\label{dt06}\rho_{\alpha\beta}^{(0)}=C^{*}_{\alpha}(0)C_{\beta}(0),\end{align}
with $\alpha,\beta=a,b,c$.

Next we seek to determine the time evolution of the density
operator. In this regard, we first assume that the three-level
atoms initially prepared in a coherent superposition of the top
and bottom levels are injected into a cavity at constant rate
$r_{a}$ and removed after some time $\tau$, which is long enough
for the atoms to decay spontaneously to levels other than the
middle or the lower. For convenience, the atomic spontaneous decay
rate $\gamma$ is taken to be the same for the two upper levels. In
the good cavity limit, $\gamma\gg\kappa$, where $\kappa$ is the
cavity damping constant, the cavity mode variables change slowly
compared with the atomic variables. Hence the atomic variables
will reach steady state in relatively short time. In this case,
the time derivative of such variables can be set to zero, while
keeping the remaining terms at time $t$. This procedure is usually
referred to as the adiabatic approximation scheme. Moreover, since
the coupling constant is believed to be small, we confine
ourselves to a linear analysis that amounts to dropping the higher
order terms in $g$.

 Now applying the linear and adiabatic approximation
schemes, it can be established in the good cavity limit  that the
time evolution of the density operator for the cavity mode, driven
by a coherent light on resonance, and coupled to a vacuum
reservoir, takes the form
\begin{align}\label{dt07}\frac{d\hat{\rho}(t)}{dt} &=
 {AC\over2B}\big[2\hat{a}^{\dagger}\hat{\rho}\hat{a} -
\hat{a}\hat{a}^{\dagger}\hat{\rho} -
\hat{\rho}\hat{a}\hat{a}^{\dagger}\big]\notag\\&+{AD\over
2B}\big[2\hat{a}\hat{\rho}\hat{a}^{\dagger} -
\hat{a}^{\dagger}\hat{a}\hat{\rho} -
\hat{\rho}\hat{a}^{\dagger}\hat{a}\big]\notag\\& +
\frac{AE}{2B}\big[\hat{a}^{\dagger}\hat{\rho}\hat{a}^{\dagger}-\hat{a}^{2}\hat{\rho}
-
\hat{\rho}\hat{a}^{\dagger^{2}}+\hat{a}\hat{\rho}\hat{a}\big]\notag\\&
+\frac{AF}{2B}\big[\hat{a}^{\dagger}\hat{\rho}\hat{a}^{\dagger}
-\hat{a}^{\dagger^{2}}\hat{\rho}- \hat{\rho}\hat{a}^{2}+
\hat{a}\hat{\rho}\hat{a}\big],\end{align} where
\begin{align}\label{dt08}A
=\frac{2r_{a}g^{2}}{\gamma^{2}},\end{align} is the linear gain
coefficient,
\begin{align}\label{dt09}B=\left(1+{\Omega^{2}\over\gamma^{2}}\right)
\left(1+{\Omega^{2}\over4\gamma^{2}}\right),\end{align}

\begin{align}\label{dt10}C= \rho_{aa}^{(0)}\left(1+{\Omega^{2}\over4\gamma^{2}}\right)-\rho_{ac}^{(0)}{3\Omega\over2\gamma} +
\rho_{cc}^{(0)}{3\Omega^{2}\over4\gamma^{2}},\end{align}

\begin{align}\label{dt11}D=\rho_{aa}^{(0)}{3\Omega^{2}\over4\gamma^{2}}+\rho_{ac}^{(0)}{3\Omega\over2\gamma}
+\rho_{cc}^{(0)}\left(1+{\Omega^{2}\over4\gamma^{2}}\right),\end{align}

 \begin{align}\label{dt12}E&=-\rho_{aa}^{(0)}{\Omega\over2\gamma}
 \left(1-{\Omega^{2}\over2\gamma^{2}}\right)-
 \rho_{ac}^{(0)}\left(1-{\Omega^{2}\over2\gamma^{2}}\right)\notag\\&+\rho_{cc}^{(0)}{\Omega\over\gamma}\left(
 1+{\Omega^{2}\over4\gamma^{2}}\right),\end{align}

\begin{align}\label{dt13}F&=-\rho_{aa}^{(0)}{\Omega\over\gamma}
\left(1+{\Omega^{2}\over4\gamma^{2}}\right)-
\rho_{ac}^{(0)}\left(1-{\Omega^{2}\over2\gamma^{2}}\right)\notag\\&+\rho_{cc}^{(0)}{\Omega\over2\gamma}\left(1-{\Omega^{2}\over2\gamma^{2}}\right).\end{align}

On the other hand, the time evolution of the density operator for
a single-mode cavity radiation coupled to a vacuum reservoir via a
single-port mirror is found using the standard method \cite{13} to
be
\begin{align}\label{dt14}\frac{d\hat{\rho}(t)}{dt}& = -i[\hat{H}_{S}(t), \;\hat{\rho}(t)] \notag\\&+
\frac{\kappa}{2} [2\hat{a}\hat{\rho}\hat{a}^{\dagger} -
\hat{a}^{\dagger}\hat{a}\hat{\rho} -
\hat{\rho}\hat{a}^{\dagger}\hat{a}],\end{align} where $\kappa$ is
the cavity damping constant. With the aid of Eqs. \eqref{dt07} and
\eqref{dt14}, the master equation describing the cavity radiation
of the driven degenerate three-level cascade laser coupled to a
vacuum reservoir turns out to be
\begin{align}\label{dt15}\frac{d\hat{\rho}(t)}{dt} &=
 {AC\over2B}\big[2\hat{a}^{\dagger}\hat{\rho}\hat{a} -
\hat{a}\hat{a}^{\dagger}\hat{\rho} -
\hat{\rho}\hat{a}\hat{a}^{\dagger}\big]\notag\\&+{1\over2}\left({AD\over
B}+\kappa\right)\big[2\hat{a}\hat{\rho}\hat{a}^{\dagger} -
\hat{a}^{\dagger}\hat{a}\hat{\rho} -
\hat{\rho}\hat{a}^{\dagger}\hat{a}\big]\notag\\& +
\frac{AE}{2B}\big[\hat{a}^{\dagger}\hat{\rho}\hat{a}^{\dagger}-\hat{a}^{2}\hat{\rho}
-
\hat{\rho}\hat{a}^{\dagger^{2}}+\hat{a}\hat{\rho}\hat{a}\big]\notag\\&
+\frac{AF}{2B}\big[\hat{a}^{\dagger}\hat{\rho}\hat{a}^{\dagger}
-\hat{a}^{\dagger^{2}}\hat{\rho}- \hat{\rho}\hat{a}^{2}+
\hat{a}\hat{\rho}\hat{a}\big].\end{align}

\section{Stochastic differential equation}

We now proceed to drive the pertinent stochastic differential
equations associated with the normal ordering. To this end, making
use of Eq. \eqref{dt15} and the fact that
\begin{align}\label{dt16}{d\over dt}\langle\hat{a}(t)\rangle =
Tr\left({d\hat{\rho}\over dt}\hat{a}\right),\end{align} one can
readily see that
\begin{align}\label{dt17}\frac{d}{dt}\langle\hat{a}(t)\rangle = -\frac{\mu}{2}\langle\hat{a}(t)\rangle
+ \beta\langle\hat{a}^{\dagger}(t)\rangle,\end{align}
\begin{align}\label{dt18}\frac{d}{dt}\langle\hat{a}^{2}(t)\rangle = -\mu\langle\hat{a}^{2}(t)\rangle +
2\beta\langle\hat{a}^{\dagger}(t)\hat{a}(t)\rangle-{AF\over B}
,\end{align}
\begin{align}\label{dt19}\frac{d}{dt}\langle\hat{a}^{\dagger}(t)\hat{a}(t)\rangle &=
-\mu\langle\hat{a}^{\dagger}(t)\hat{a}(t)\rangle \notag\\&+
\beta\big[\langle\hat{a}^{\dagger^{2}}(t)\rangle +
\langle\hat{a}^{2}(t)\rangle\big] + {AC\over B},\end{align} where
\begin{align}\label{dt20}\mu = {A\over B}(D-C)+\kappa,\end{align}
\begin{align}\label{dt21}\beta = {A\over2B}(E-F).\end{align}

We notice that the operators in Eqs. \eqref{dt17}, \eqref{dt18},
and \eqref{dt19} are in the normal order. Hence we can express
these equations in terms of the  c-number variables associated
with the normal ordering as
\begin{align}\label{dt22}\frac{d}{dt}\langle\alpha(t)\rangle = -\frac{\mu}{2}\langle\alpha(t)\rangle
+ \beta\langle\alpha^{*}(t)\rangle,\end{align}
\begin{align}\label{dt23}\frac{d}{dt}\langle\alpha^{2}(t)\rangle = -\mu\langle\alpha^{2}(t)\rangle +
2\beta\langle\alpha^{*}(t)\alpha(t)\rangle-{AF\over B}
,\end{align}
\begin{align}\label{dt24}\frac{d}{dt}\langle\alpha^{*}(t)\alpha(t)\rangle &=
-\mu\langle\alpha^{*}(t)\alpha(t)\rangle \notag\\&+
\beta\big[\langle\alpha^{*^{2}}(t)\rangle +
\langle\alpha^{2}(t)\rangle\big] + {AC\over B}.\end{align}

On the basis of Eqs. \eqref{dt22}, it is possible to write
\begin{align}\label{dt25}\frac{d}{dt}\alpha(t) = -\frac{\mu}{2}\alpha(t)
+ \beta\alpha^{*}(t) + E(t),\end{align}
 where
$E(t)$ is the corresponding noise force the properties of which
remain to be determined.

The expectation value of Eq. \eqref{dt25} has the same form as
\eqref{dt22} provided that the noise force has a zero mean,
\begin{align}\label{dt26}\langle E(t)\rangle = 0.\end{align}
One can also verify applying Eqs. \eqref{dt22}, \eqref{dt23},
\eqref{dt24}, \eqref{dt25}, along with the fact that the noise
force at time $t$ does not correlate with the cavity mode
variables at the earlier times that
\begin{align}\label{dt33}\langle E(t')E(t)\rangle =-{AF\over B}\delta(t-t'),\end{align}
\begin{align}\label{dt34}\langle E(t)E^{*}(t')\rangle = {AC\over B}\delta(t-t').\end{align}
We notice that Eqs. \eqref{dt26}, \eqref{dt33}, and \eqref{dt34}
represent the mean and correlation properties of the noise force.

Furthermore, it proves to be more convenient to introduce new
variables defined as
\begin{align}\label{dt35}\alpha_{\pm}(t) =
\alpha^{*}(t)\pm\alpha(t),\end{align}so that one can easily see
with the aid of Eq. \eqref{dt25} and its complex conjugate that
\begin{align}\label{dt36}{d\over dt}\alpha_{\pm}(t) =
-{\lambda_{\mp}\over2}\alpha_{\pm}(t) + E^{*}(t)\pm
E(t),\end{align} where
\begin{align}\label{dt37}\lambda_{\mp}=\mu\mp2\beta.\end{align}
The formal integration of Eq. \eqref{dt36} results in
\begin{align}\label{dt39}\alpha(t) = a_{+}(t)\alpha(0) + a_{-}(t)\alpha^{*}(0)+F_{-}(t)+F_{+}(t),\end{align}
in which
\begin{align}\label{dt40}a_{\pm}(t)={1\over2}\left(e^{-{\lambda_{-}t\over2}}
\pm e^{-{\lambda_{+}t\over2}}\right),\end{align}
\begin{align}\label{dt42}F_{\pm}(t)={1\over2}\int_{0}^{t}e^{-{\lambda_{\mp}t\over2}}[E(t')\pm E^{*}(t')]dt'.\end{align}
We observe that a  well-behaved solution of Eq. \eqref{dt36}
exists at steady state for $\lambda_{\mp}>0$. As a result, $\mu =
2\beta$ is designated as a threshold condition. It may worth
mentioning that the squeezing as well as the statistical
properties of the cavity radiation can be studied making use of
Eq. \eqref{dt39}.

\section{ Quadrature variance}

 We now seek to evaluate the variance of the quadrature operators that are defined
 for a single-mode cavity radiation by
 \begin{align}\label{dt43}\hat{a}_{+} = \hat{a}^{\dagger} + \hat{a}\end{align}
and
\begin{align}\label{dt44}\hat{a}_{-} = i(\hat{a}^{\dagger} - \hat{a}).\end{align}
Employing the boson commutation relation, the variance of these
quadrature operators can be expressed in terms of the
corresponding c-number variables associated with the normal
ordering as
\begin{align}\label{dt47}\Delta a^{2}_{\pm} &= 1 + 2\langle\alpha^{*}\alpha\rangle \pm
\langle\alpha^{*^{2}}\rangle \pm \langle\alpha^{2}\rangle
\notag\\&\mp \langle\alpha^{*}\rangle^{2} \mp
\langle\alpha\rangle^{2} -
2\langle\alpha^{*}\rangle\langle\alpha\rangle.\end{align}

Next we proceed to evaluate various correlations involved in Eq.
\eqref{dt47}. To begin with, for the cavity mode initially in a
vacuum state, we see from Eq. \eqref{dt39} that
\begin{align}\label{dt48}\langle\alpha_{\pm}(t)\rangle =0,\end{align}
as a result
\begin{align}\label{dt49}\Delta a^{2}_{\pm} =
1\pm\langle\alpha^{2}_{\pm}(t)\rangle.\end{align} On the other
hand, in view of Eq. \eqref{dt36} it is possible to write
\begin{align}\label{dt50}{d\over dt}\langle\alpha^{2}_{\pm}(t)\rangle &=
-\lambda_{\mp}\langle\alpha^{2}_{\pm}(t)\rangle \notag\\&+
2\langle\alpha_{\pm}(t)E^{*}(t)\rangle\pm
2\langle\alpha_{\pm}(t)E(t)\rangle.\end{align}
 Thus taking Eqs. \eqref{dt33} and \eqref{dt34} into consideration, we find for a
cavity mode initially in a vacuum state
\begin{align}\label{dt54}\langle\alpha^{2}_{\pm}(t)\rangle =
 -{2A\over B\lambda_{\mp}}\big[F\mp
C\big]\big(1-e^{-\lambda_{\mp}t}\big).\end{align}

 In order to put Eq. \eqref{dt54} in a more convenient manner, we
 express the initial atomic coherence of the top and bottom levels as
 \begin{align}\label{dt55}\rho_{ac}^{(0)} =
 |\rho_{ac}^{(0)}|e^{i\theta},\end{align} where $\theta$ is the
 phase factor. It can also be easily checked that
\begin{align}\label{dt56}|\rho_{ac}^{(0)}| = \sqrt{\rho_{aa}^{(0)}\rho_{cc}^{(0)}}.\end{align}
Moreover, upon introducing a new parameter defined by
\begin{align}\label{dt57}\rho_{aa}^{(0)}={1-\eta\over2},\end{align}
with $-1\le\eta\le1$, we  see
that
\begin{align}\label{dt58}\rho_{cc}^{(0)} =
\frac{1+\eta}{2},\end{align}
\begin{align}\label{dt59}\rho_{ac}^{(0)} = \frac{\sqrt{1-\eta^{2}}}{2}e^{i\theta}.\end{align}

Therefore, on the basis of Eqs. \eqref{dt09}, \eqref{dt10},
\eqref{dt11}, \eqref{dt12}, \eqref{dt13}, \eqref{dt20},
\eqref{dt21}, \eqref{dt37}, \eqref{dt54}, \eqref{dt57}, and
\eqref{dt58}, Eq. \eqref{dt59} finally takes, for $\theta=0$,  the
form
\begin{align}\label{dt60}\langle\alpha^{2}_{\pm}(t)\rangle
&=\left\{
 {A\left[{\Omega\over2\gamma}\left(1-3\eta+
{\Omega^{2}\over\gamma^{2}}\right)+\sqrt{1-\eta^{2}}\left(1-{\Omega^{2}\over2\gamma^{2}}\right)
\right]\over\chi_{\pm}}\right.\notag\\&\left.\pm
{A\left[1-\eta+{\Omega^{2}\over2\gamma^{2}}(2+\eta)-\sqrt{1-\eta^{2}}{3\Omega\over2\gamma}\right]\over\chi_{\pm}}
\right\}\notag\\&\times\big(1-e^{-\lambda_{\mp}t}\big),\end{align}
which reduces at steady state to
\begin{align}\label{dt61}\langle\alpha^{2}_{\pm}(t)\rangle_{ss} &=
 {A\left[{\Omega\over2\gamma}\left(1-3\eta+
{\Omega^{2}\over\gamma^{2}}\right)+\sqrt{1-\eta^{2}}\left(1-{\Omega^{2}\over2\gamma^{2}}\right)\right]\over\chi_{\pm}}\notag\\&\pm
{A\left[1-\eta+{\Omega^{2}\over2\gamma^{2}}(2+\eta)-\sqrt{1-\eta^{2}}{3\Omega\over2\gamma}\right]\over\chi_{\pm}}
,\end{align}
 in which
\begin{align}\label{dt62}\chi_{\pm}&=\kappa\left(1+{\Omega^{2}\over\gamma^{2}}\right)\left(1+{\Omega^{2}\over4\gamma^{2}}\right)
+A\left[\left(1-{\Omega^{2}\over2\gamma^{2}}\right)\eta\right.\notag\\&\left.
+\sqrt{1-\eta^{2}}{3\Omega\over2\gamma}\mp{\Omega\over2\gamma}
\left(1+{\Omega^{2}\over\gamma^{2}}\right)\right].\end{align}
 Hence
the variance of the quadrature operators \eqref{dt49} at steady
state turn out to be
\begin{align}\label{dt63}\Delta a^{2}_{\pm} &= 1\pm{A\left[{\Omega\over2\gamma}\left(1-3\eta+
{\Omega^{2}\over\gamma^{2}}\right)+
\sqrt{1-\eta^{2}}\left(1-{\Omega^{2}\over2\gamma^{2}}\right)\right]\over\chi_{\pm}}\notag\\&+
{A\left[1-\eta+{\Omega^{2}\over2\gamma^{2}}(2+\eta)-\sqrt{1-\eta^{2}}{3\Omega\over2\gamma}\right]\over\chi_{\pm}}.\end{align}

\begin{figure}[hbt]
\centerline{\includegraphics [height=6.5cm,angle=0]{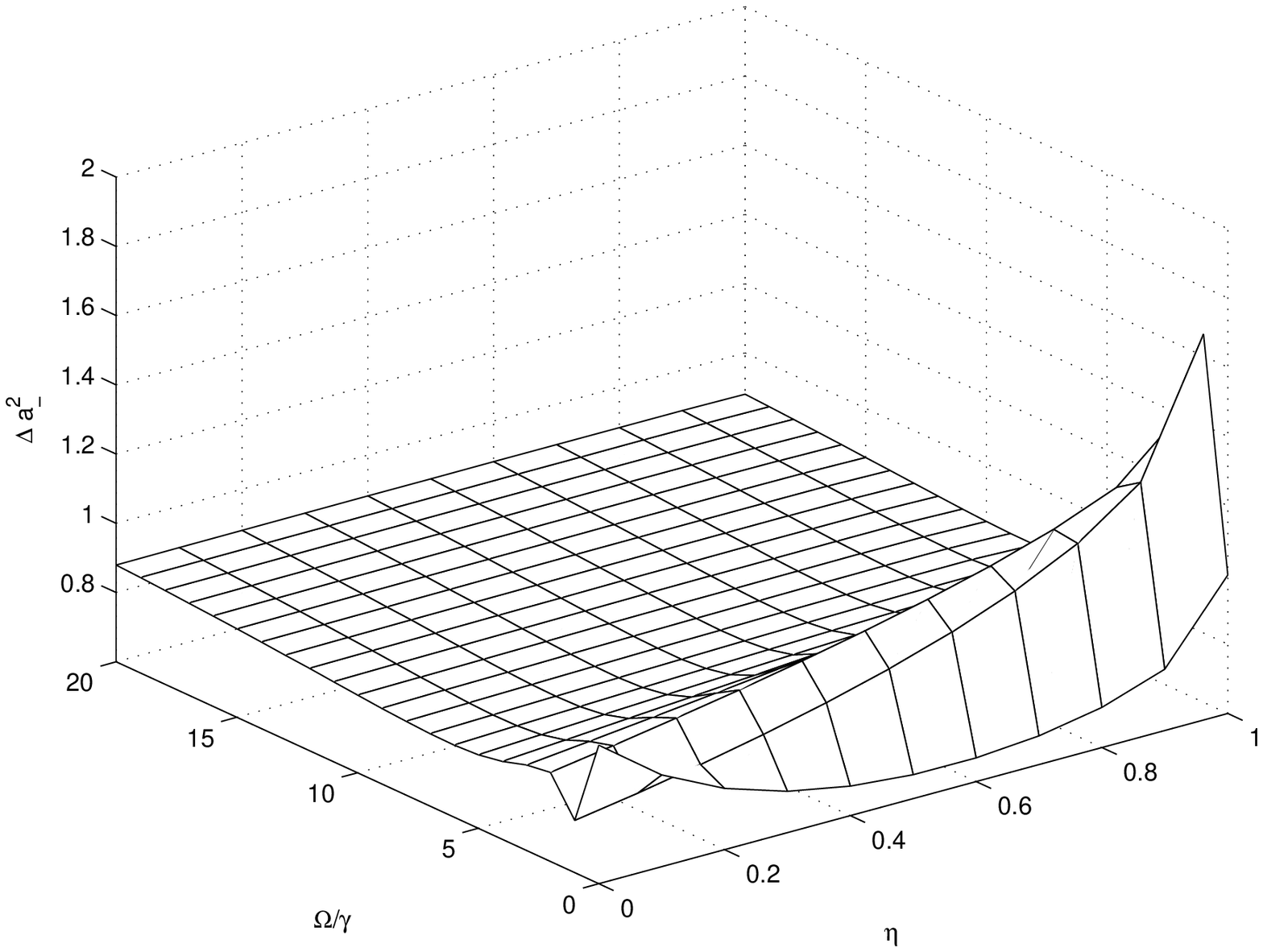} }
\footnotesize {FIG. 2: Plot of the quadrature variance $\Delta
a_{-}^{2}$ for the cavity radiation at steady state
 for $\kappa = 0.2$, $\theta=0$, and $A=0.33$.}
\end{figure}

We see from Fig. (2) that the light produced by a degenerate
three-level cascade laser externally driven on resonance by a
coherent radiation exhibits squeezing for certain values of
$\Omega/\gamma$ and $\eta$ for a given linear gain coefficient. It
is found for $\kappa=0.2$ and $A=0.33$ that the squeezing occurs
for all values of $\Omega/\gamma$ when $\eta<0.5$. It is also
possible to notice that the more there are atoms initially in the
upper level, the better would be the resulting squeezing.

In order to study the dependence of the squeezing on the amplitude
of the driving radiation and initially injected atomic coherence
closely we consider various cases of interest. In this respect, it
is not difficult to check for $\Omega =0$ that
\begin{align}\label{dt64}\Delta a^{2}_{\pm} = {\kappa+A(1\pm\sqrt{1-\eta^{2}})\over A\eta
+\kappa}.\end{align} The same result has been obtained for
instance by Fesseha \cite{11}.

\begin{figure}[hbt]
\centerline{\includegraphics [height=6.5cm,angle=0]{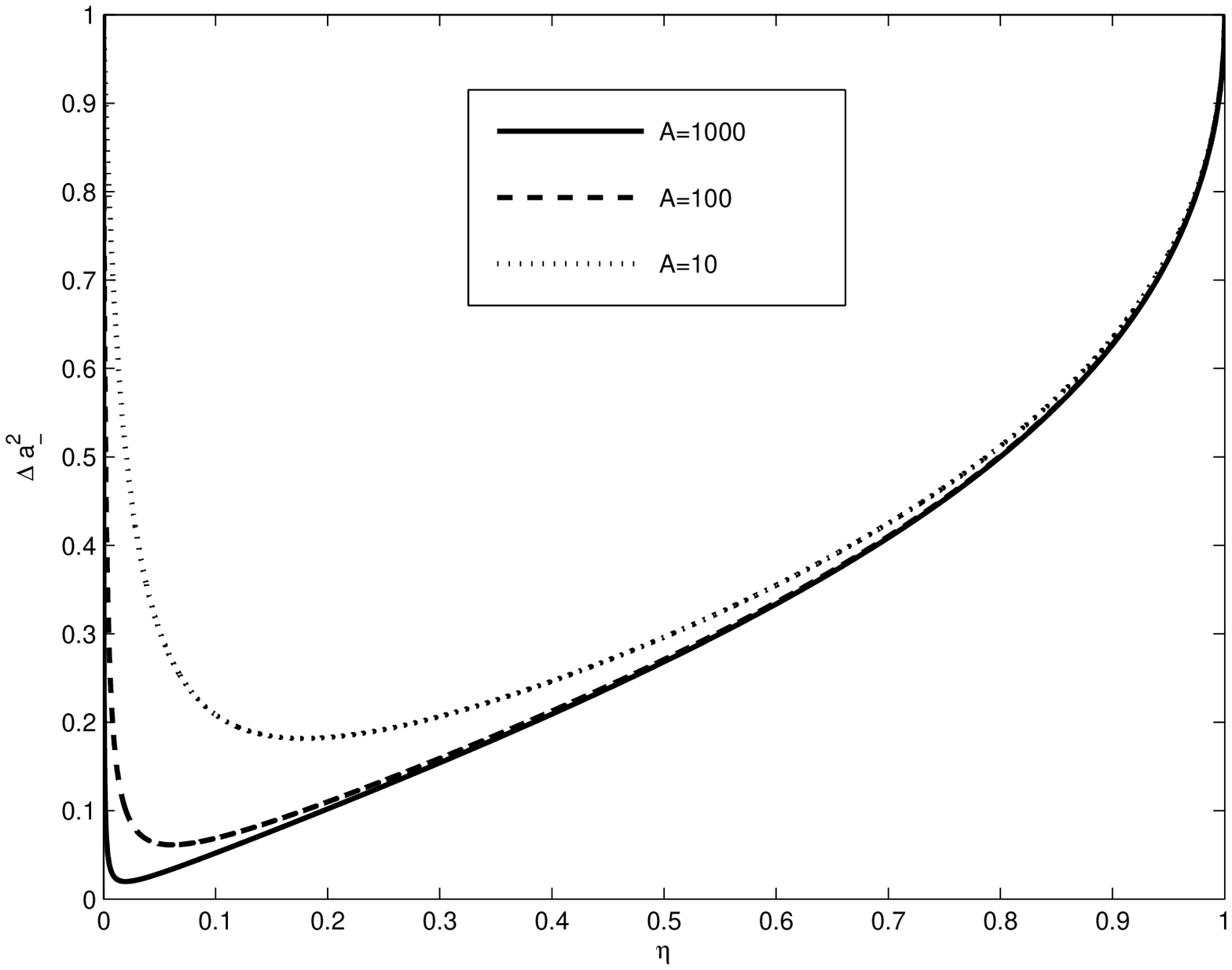}}
FIG. 3:~\footnotesize { Plots of the quadrature variance $\Delta
a_{-}^{2}$ of the cavity radiation at steady state
 for $\kappa = 0.2$, $\theta=0$, $\Omega=0$, and different values of
 $A$.}
\end{figure}
We clearly see from Fig. (3) that the degree of squeezing
increases with the linear gain coefficient and a substantial
degree of squeezing is found for small values of $\eta$. This
indicates that the more atoms are injected into the cavity at a
time the more the degree of the squeezing of the cavity radiation
would be. In particular, a maximum of 98\% squeezing occurs at
$\eta=0.02$ for $A=1000$. The correlated emission initiated by the
initial atomic coherence is responsible for the reduction of the
fluctuations of the noise in one of the quadrature components
below the classical limit.

Moreover, if initially all atoms are in the lower level, $\eta=1$,
Eq. \eqref{dt63} takes the form
\begin{align}\label{dt65}\Delta a^{2}_{\pm} = 1\mp
{A\left[{\Omega\over\gamma}-
{\Omega^{3}\over2\gamma^{3}}\mp{3\Omega^{2}\over2\gamma^{2}}
\right]\over\chi'_{\pm}},\end{align} with
\begin{align}\label{dt66}\chi'_{\pm}&=\kappa\left(1+{\Omega^{2}\over\gamma^{2}}\right)\left(1+{\Omega^{2}\over4\gamma^{2}}\right)
+A\left[1-{\Omega^{2}\over2\gamma^{2}}\right.\notag\\&\left.
\mp{\Omega\over2\gamma}
\left(1+{\Omega^{2}\over\gamma^{2}}\right)\right].\end{align}
\begin{figure}[hbt]
\centerline{\includegraphics [height=6.5cm,angle=0]{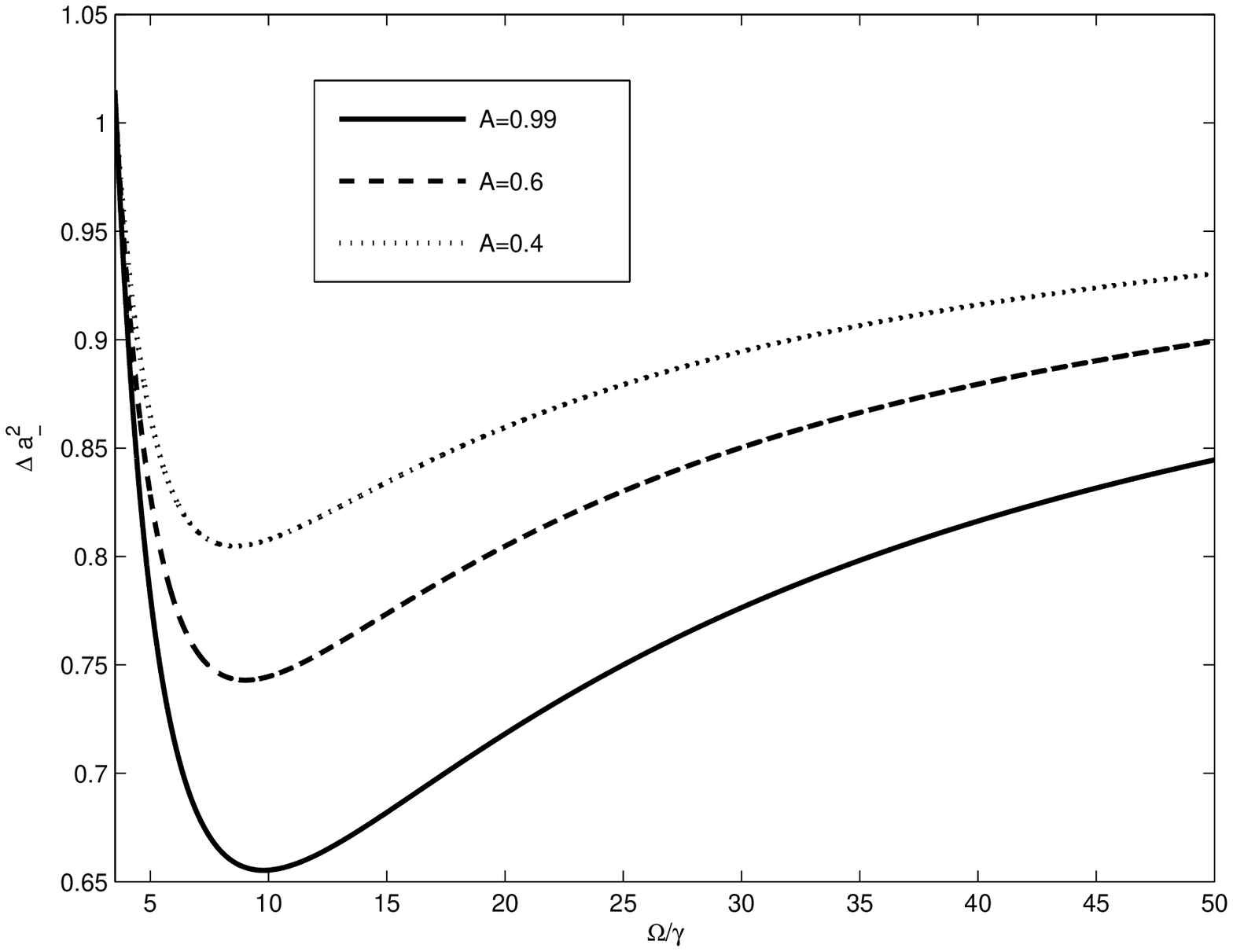}
} FIG. 4:~\footnotesize { Plots of the quadrature variance $\Delta
a_{-}^{2}$ of the cavity radiation at steady state
 for $\kappa = 0.2$, $\theta=0$, $\eta=1$, and different values of $A$.}
\end{figure}
We found that the light generated when all atoms are initially in
the lower level exhibits squeezing for $\Omega>3.5\gamma$. In this
case a significant squeezing is obtained in the vicinity of a
particular amplitude of the driving radiation for each value of
the linear gain coefficient. It is also found  that a squeezing of
nearly 35\% occurs at $\Omega=10.1\gamma$ for $A=0.99$. In
addition, it is not difficult to see from Fig. (4) that the degree
of squeezing decreases with the amplitude of the driving radiation
for larger values of $\Omega/\gamma$. Though the squeezing
increases with the linear gain coefficient in this case, we cannot
use arbitrary values of $A$, since the steady state consideration
fails to be applied for $A>0.99$, for $\eta=1$ and
$\Omega=3.5\gamma$.

Furthermore, when the atoms are initially prepared with equal
probability to be in the top and bottom levels, $\eta=0$, we get
\begin{align}\label{dt67}\Delta a^{2}_{\pm}(t) &= 1\pm{A\left[{\Omega\over2\gamma}\left(
{\Omega^{2}\over2\gamma^{2}}-2-{\Omega\over\gamma}\right)+1\right]\over\chi''_{\pm}}\notag\\&
+{A\left[
1+{\Omega^{2}\over\gamma^{2}}-{3\Omega\over2\gamma}\right]\over\chi''_{\pm}}
,\end{align}
 in which
\begin{align}\label{dt68}\chi''_{\pm}&=\kappa\left(1+{\Omega^{2}\over\gamma^{2}}\right)\left(1+{\Omega^{2}\over4\gamma^{2}}\right)
\notag\\&+ A\left[{3\Omega\over2\gamma}\mp {\Omega\over2\gamma}
\left(1+{\Omega^{2}\over\gamma^{2}}\right)\right].\end{align}

\begin{figure}[hbt]
\centerline{\includegraphics [height=6.5cm,angle=0]{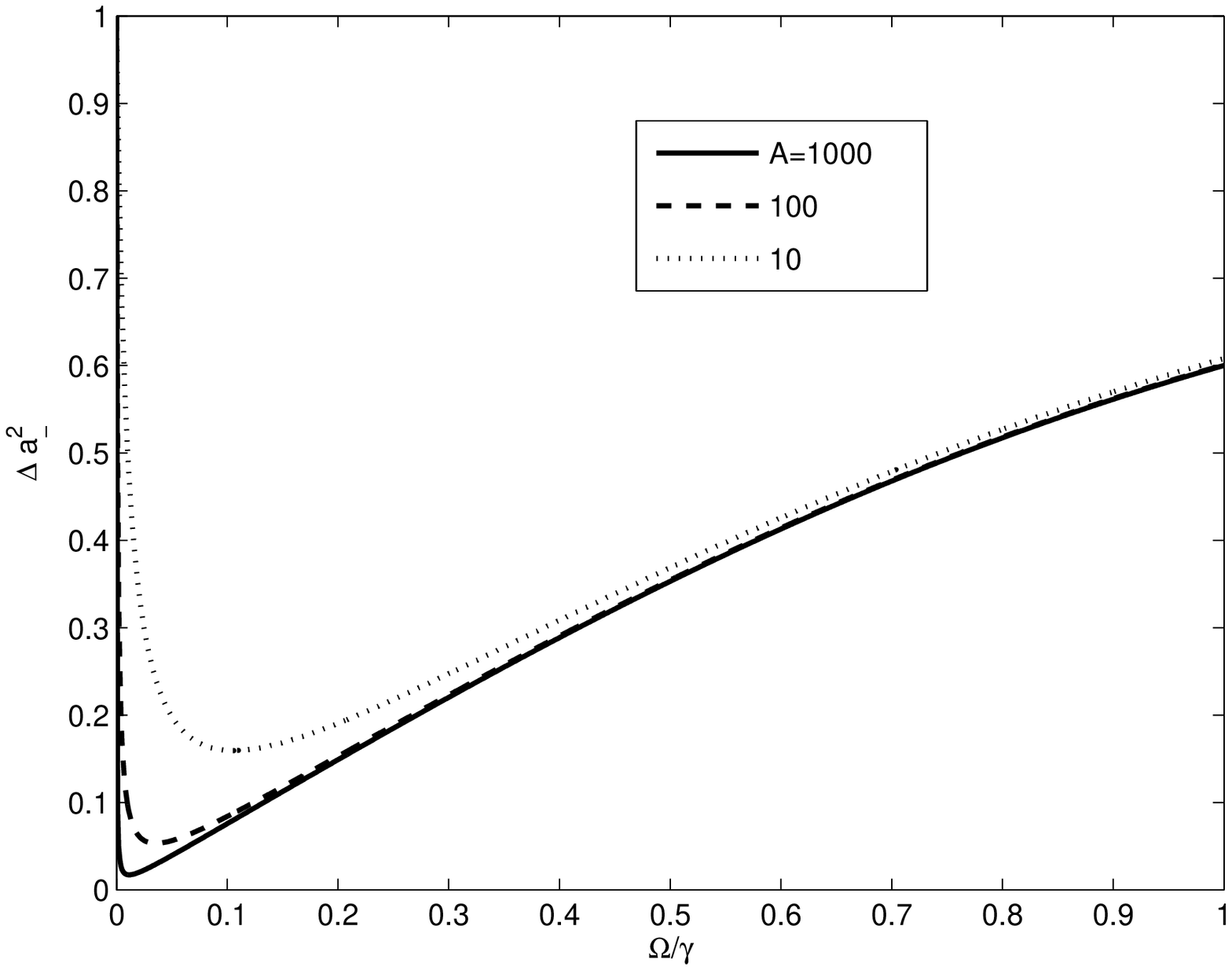}
} FIG. 5:~\footnotesize { Plots of the quadrature variance $\Delta
a_{-}^{2}$ of the cavity radiation at steady state
 for $\kappa = 0.2$, $\theta=0$, $\eta=0$, and different values of $A$.}
\end{figure}
We see from Fig. (5) that the light produced by the degenerate
three-level laser, when the atoms are initially prepared with
equal probability of being in the top and bottom levels, exhibits
substantial degree of squeezing for small values of
$\Omega/\gamma$. It is found that a maximum squeezing of 98.3\%
occurs at $\Omega=0.012\gamma$ for $A=1000$. As we have seen
before the degree of squeezing decreases with the amplitude of the
classical radiation for larger values of $\Omega/\gamma$, but it
increases with the linear gain coefficient throughout.

On the basis of the definition of the parameter $\eta$ (Eq.
\eqref{dt57}) we notice that for $\eta=0$,
$\rho_{aa}^{(0)}=\rho_{cc}^{(0)}=\rho_{ac}^{(0)}=1/2$, which
corresponds to a maximum initial atomic coherence. But, for
$\eta=1$, $\rho_{aa}^{(0)}=\rho_{ac}^{(0)}=0$ and $
\rho_{cc}^{(0)}=1$, which is related to the absence of injected
atomic coherence at the beginning. It is not difficult to see from
Eq. \eqref{dt64} that there is no squeezing property when the
atoms are initially prepared with maximum or minimum atomic
coherence, if they are not driven externally ($\Omega=0$).
However, as shown in Fig. (3) the maximum squeezing occurs when
the atoms are prepared with initial coherence very close to the
maximum possible value in this case. We also observe from Figs.
(4) and (5) that the external coherent radiation initiates the
correlation between the photons which leads to  squeezing when
$\eta=0$ or $\eta=1$. When there is no injected atomic coherence a
squeezing close to 50\% is obtained near particular amplitude of
the external radiation. This result agrees with the prediction of
H. Xiong et al \cite{6} that three-level laser in which the atoms
are initially prepared in the bottom level and externally driven
by strong radiation resembles parametric oscillator.

 In addition, we realize upon comparing the
results shown in Figs. (3), (4), and (5) that a better squeezing
can be obtained when the atoms are initially prepared with a
maximum atomic coherence and also driven externally with a
coherent radiation of relatively small amplitude. Likewise, the
maximum noise reduction when the atoms are injected into a
resonant cavity with maximum atomic coherence and driven with
classical radiation for $\Lambda$ three-level laser is obtained
\cite{1}. Though the external radiation induces the coherence
which is believed to be the cause of squeezing, we observe that
pumping the atoms with stronger radiation than required destroys
squeezing. In connection to what we have seen before, we note that
the degree of squeezing would be maximum  for certain values of
the initial atomic coherence that depends on the rate at which the
atoms are injected into the cavity.

We hence conclude that the cavity radiation exhibits significant
squeezing for certain values of the amplitude of the driving
radiation and initial preparation of the superposition, where the
degree of squeezing increases with the linear gain coefficient.

\section{ Mean photon number}

The mean photon number of the cavity radiation
\begin{align}\label{dt69}\bar{n}=\langle\alpha^{*}(t)\alpha(t)\rangle\end{align}
 can be expressed,
with the aid of Eq. \eqref{dt35}, in the form
\begin{align}\label{dt70}\bar{n} = {\langle\alpha^{2}_{+}(t)\rangle -
\langle\alpha^{2}_{-}(t)\rangle\over4}.\end{align} On account of
Eq. \eqref{dt61}, we then find at steady state
\begin{align}\label{dt71}\bar{n}&=-{A\left[{\Omega\over2\gamma}(1-3\eta)+
{\Omega^{3}\over2\gamma^{3}}-
\left(1-\eta+{\Omega^{2}\over2\gamma^{2}}(2+\eta)\right)\right]\over4\chi_{-}}\notag\\&
+{A\sqrt{1-\eta^{2}}\left({\Omega^{2}\over2\gamma^{2}}-1-
{3\Omega\over2\gamma} \right)\over4\chi_{-}}
\notag\\&+{A\left[{\Omega\over2\gamma}(1-3\eta)+
{\Omega^{3}\over2\gamma^{3}}+
\left(1-\eta+{\Omega^{2}\over2\gamma^{2}}(2+\eta)\right)\right]\over4\chi_{+}}\notag\\&
-{A\sqrt{1-\eta^{2}}\left({\Omega^{2}\over2\gamma^{2}}-1+
{3\Omega\over2\gamma} \right)\over4\chi_{+}}.\end{align}

\begin{figure}[hbt]
\centerline{\includegraphics [height=6.5cm,angle=0]{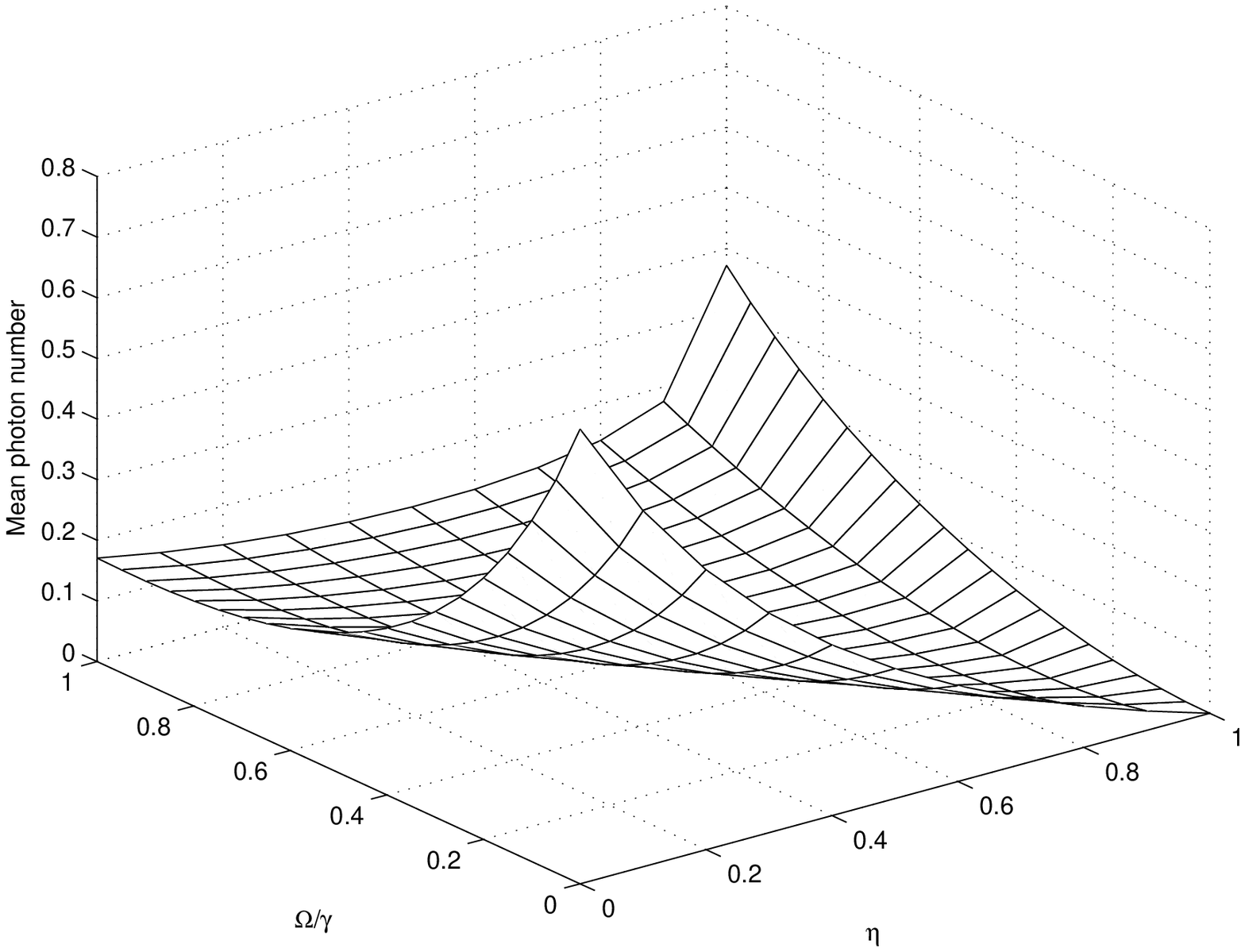} }
\footnotesize {FIG. 6: Plot of the mean photon number of the
cavity radiation at steady state
 for $\kappa = 0.2$, $\theta=0$, and $A=0.3$.}
\end{figure}
One can clearly see from Fig. (6) that no light is produced when
initially all atoms are in the bottom level and if there is no
external driving radiation. We also notice that as a result of the
external driving it is possible to generate an intense light from
the laser even when the atoms are initially in the lower level.
This demonstrates the mechanism of lasing without population
inversion.

Now we seek to consider various cases of interest. For $\Omega
=0$, Eq. \eqref{dt71} reduces
\begin{align}\label{dt72}\bar{n} = {A(1-\eta)\over2(A\eta+\kappa)},\end{align}
which is the same as the result obtained for instance by Fesseha
\cite{10} in the absence of the driving radiation. We see from Eq.
\eqref{dt72} that the mean photon number would be zero when there
is no driving light and all atoms are initially in the bottom
level, and one gets the most intense light when all atoms are
initially in the upper level as expected. In addition, for $\eta
=1$, Eq. \eqref{dt71} takes the form
\begin{align}\label{dt73}\bar{n}&=-{A\left[-{\Omega\over\gamma}+
{\Omega^{3}\over2\gamma^{3}}-{3\Omega^{2}\over2\gamma^{2}}
\right]\over4\chi'_{-}}\notag\\&+{A\left[-{\Omega\over\gamma}+
{\Omega^{3}\over2\gamma^{3}}+{3\Omega^{2}\over2\gamma^{2}}
\right]\over4\chi'_{+}}.\end{align}

\begin{figure}[hbt]
\centerline{\includegraphics [height=6.5cm,angle=0]{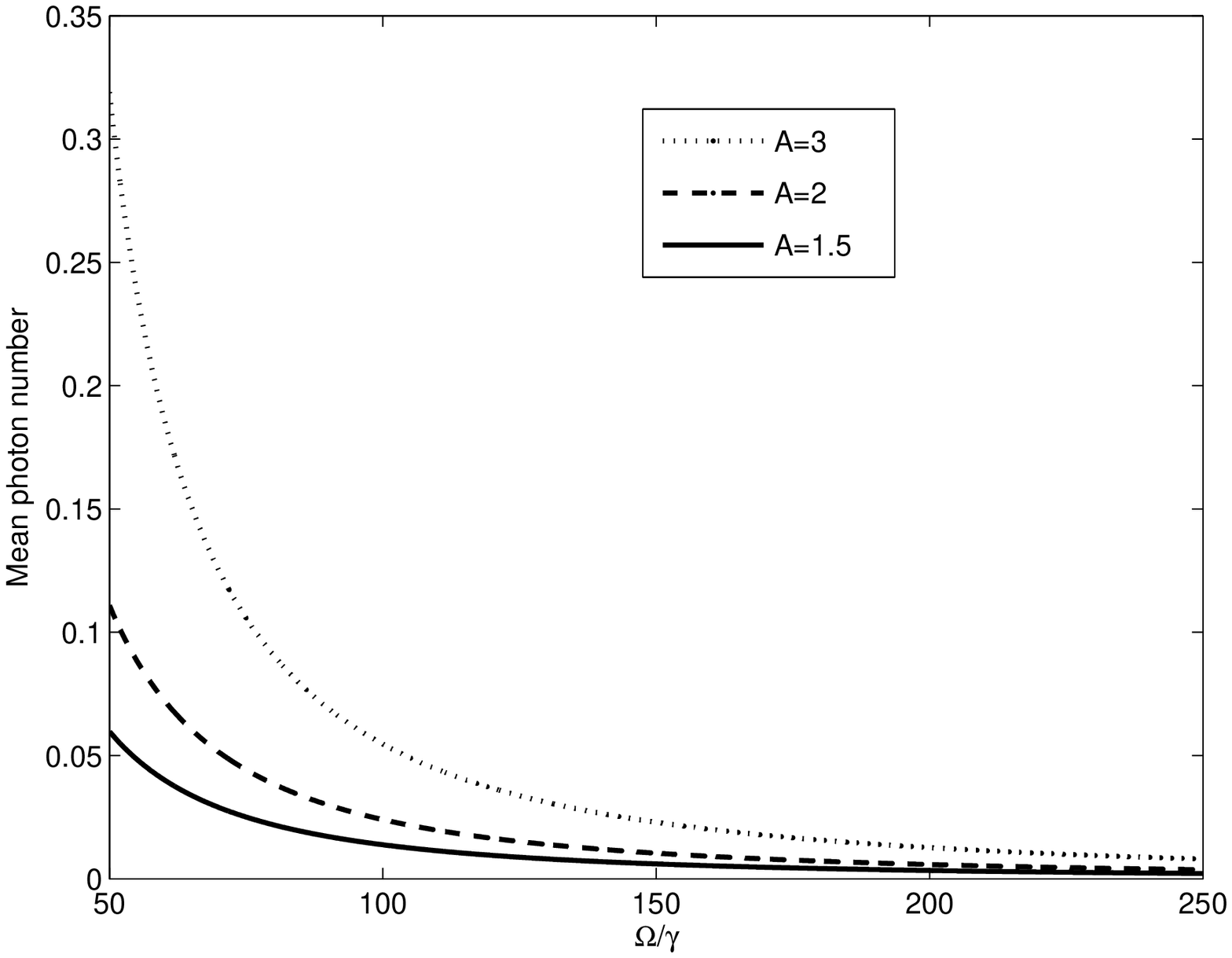} }
\footnotesize {FIG. 7: Plots of the mean photon number of the
cavity radiation at steady state
 for $\kappa = 0.2$, $\theta=0$, $\eta=1$, and different values of $A$.}
\end{figure}
As indicated in Fig. (7), the intensity of the light subsequently
decreases if we keep on increasing the strength of the driving
light.

 On the other hand, for $\eta=0$, we readily get from Eq.
 \eqref{dt71} that
\begin{align}\label{dt74}\bar{n}&=-{A\left[{\Omega\over2\gamma}\left(1+
{\Omega^{2}\over\gamma^{2}}+{\Omega\over\gamma}\right)-
{3\Omega\over2\gamma}+
{\Omega^{2}\over\gamma^{2}}\right]\over4\chi''_{-}}\notag\\&+{A\left[{\Omega\over2\gamma}\left(1+
{\Omega^{2}\over\gamma^{2}}+{\Omega\over\gamma}\right)-2-\left(
{3\Omega\over2\gamma}+
{\Omega^{2}\over\gamma^{2}}\right)\right]\over4\chi''_{+}}.\end{align}
\begin{figure}[hbt]
\centerline{\includegraphics [height=6.5cm,angle=0]{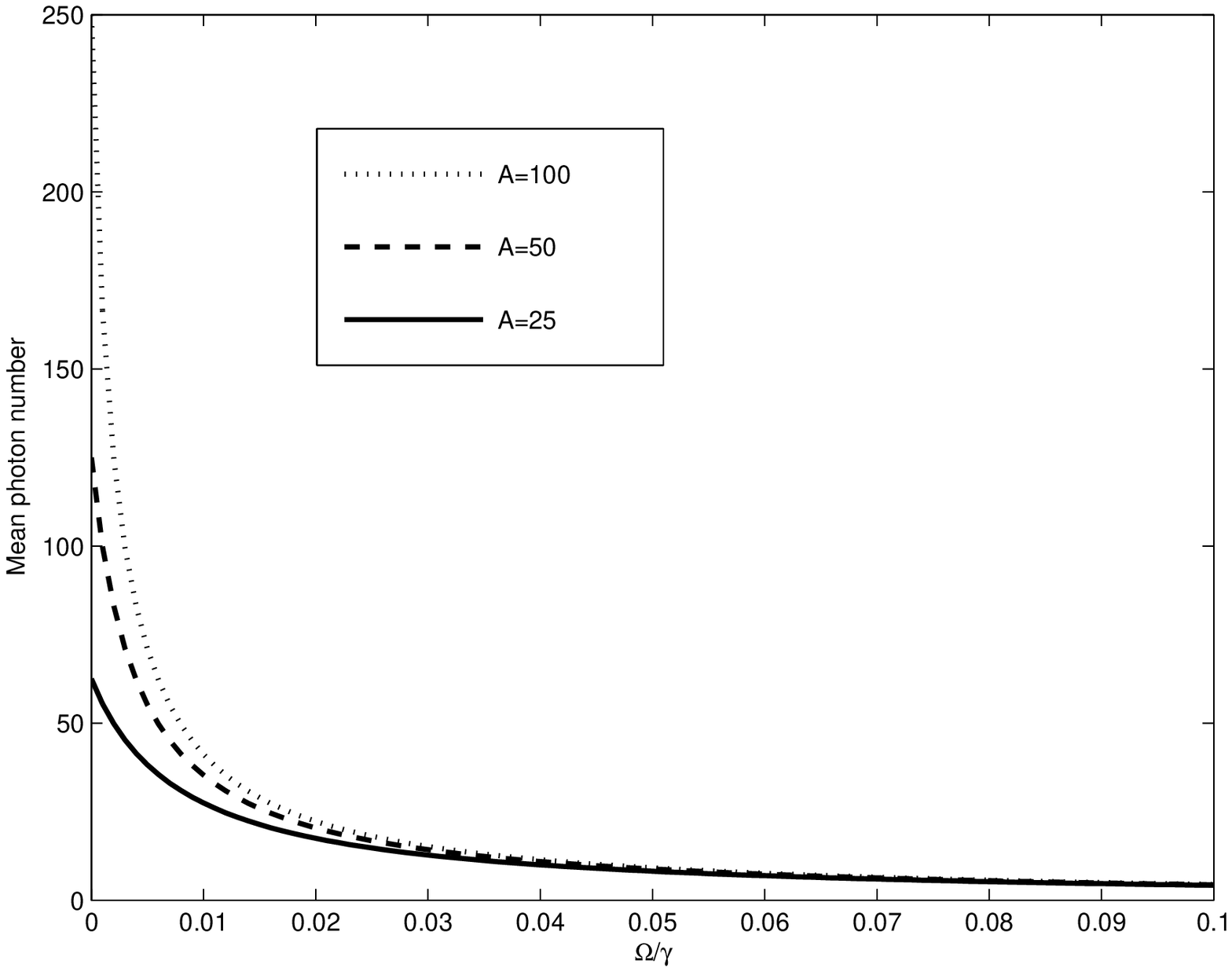} }
FIG. 8:~\footnotesize { Plots of the mean photon number of the
cavity radiation at steady state
 for $\kappa = 0.2$, $\theta=0$, $\eta=0$, and different values of $A$.}
\end{figure}

 We notice that the intensity of the produced light decreases
with the strength of the external coherent light if the atoms are
initially prepared to have equal probability of being in the top
and bottom levels for $\Omega<\gamma$.  We clearly see from Figs.
(7) and (8) that the mean photon number increases with the linear
gain coefficient.

\section{CONCLUSION}

We present a detailed analysis of the squeezing properties of the
light produced by the degenerate three-level cascade laser coupled
to a vacuum reservoir via one of the coupler mirrors and an
external resonant coherent radiation in the other. We found that
the cavity radiation exhibits up to 98.3\% squeezing under certain
conditions pertaining to the initial preparation of the
superposition and strength of the coherent radiation.

Driving the atoms with an external coherent radiation affects both
the degree of squeezing and intensity of the generated light. We
found that when the atoms are driven externally with a strong
radiation the resulting squeezing and intensity of the cavity
radiation are considerably reduced. Thus we cannot see the
practical advantageous of this mechanism in this respect. However,
we also found that an intense radiation with a substantial degree
of squeezing can be generated specially when the atoms are
initially prepared with equal probability of being in the bottom
and top levels where there is no squeezing in the absence of the
driving radiation. In addition, it is possible to get a squeezed
light when the atoms are initially in the bottom level by this
mechanism where there is no radiation at all in the absence of
driving. We hence realize that driving mechanism can be considered
as an option for producing a squeezed light when it is difficult
to prepare the atoms in an arbitrary initial superposition. On the
other hand, though it appears reasonable to expect enhancement of
squeezing when we externally drive atoms with an arbitrary initial
superposition, we are unable to confirm this for all possible
cases from our analysis. However, driving the atoms with external
radiation is found to significantly improve the squeezing, if the
atoms are prepared initially with maximum atomic coherence.


\begin{thebibliography}{1}
%%%%
\bibitem{1} C. Saaverda, J. C. Retamal, and C. H. Keitel, Phys. Rev. {\bf{A
55}}, 3802 (1997).

\bibitem{2} M. A. G. Martinez, P. R. Herczfeld, C. Samuels, L. M.
Narducci, and C. H. Keitel, Phys. Rev. {\bf{A 55}}, {\indent{4483
(1997)}}.

\bibitem{3} Y. Zhu, Phys. Rev. {\bf{A 55}}, 4568 (1997).

\bibitem{4} N. A. Ansari, J. G. Banacloche, and M. S. Zubairy, Phys. Rev.
{\bf{A 41}}, 5179 (1990).

\bibitem{5} S. An and M. Sargent III, Phys. Rev. {\bf{A 39}}, 1841 (1989)
\bibitem{6} H. Xiong, M. O. Scully, and M. S. Zubairy, Phys. Rev.
Lett. {\bf{94}}, 023601 (2005).

\bibitem{7} N. A. Ansari, Phys. Rev. {\bf{A 46}}, 1560 (1992).

\bibitem{8} M. O. Scully, K. Wodkiewicz, M. S. Zubairy, J. Bergou, N. Lu, and J. Meyer ter Van, Phys. Rev. lett.
{\bf{60}}, 1832 (1988).

\bibitem{9} J. Anwar and M. S. Zubairy, Phys. Rev. {\bf{A 49}}, 481
(1994).

\bibitem{10} N. A. Ansari, Phys. Rev. {\bf{A 48}}, 4686 (1993)

\bibitem{11} K. Fesseha, Phys. Rev. {\bf{A 63}}, 033811 (2001)
\bibitem{12} Xiang-ming Hu and Zhi-zhan Xu, J. Phys. B: At. Mol. Opt. Phys. {\bf{34}}, 787 (2001)
\bibitem{13} W. H. Louisell, {\it{Quantum statistical properties of radiation}} (Wiley, Newyork,
1973).

\end{thebibliography}
\end{document}